\documentclass[twocolumn]{aastex61}
\usepackage{color}

\newcommand       \msun        	{$M_{\odot}$}

\newcommand       \kms        {km~s$^{-1}$}

\newcommand        \mic        	 {$\mu$m}

\begin{document}

\title{	THE EVOLUTION OF DUST OPACITY IN CORE COLLAPSE SUPERNOVAE AND THE RAPID FORMATION OF DUST IN THEIR EJECTA}

\author{Eli Dwek}
\affiliation{Observational Cosmology Lab, NASA Goddard Space Flight Center, Mail Code 665, Greenbelt, MD 20771, USA}
\email{eli.dwek@nasa.gov}

\author{Arkaprabha Sarangi}
\affiliation{CRESST~II/CUA/GSFC}
\affiliation{Observational Cosmology Lab, NASA Goddard Space Flight Center, Mail Code 665, Greenbelt, MD 20771, USA}

\author{Richard G. Arendt}
\affiliation{CRESST~II/UMBC/GSFC}
\affiliation{Observational Cosmology Lab, NASA Goddard Space Flight Center, Mail Code 665, Greenbelt, MD 20771, USA}

\received{receipt date}
\revised{revision date}
\accepted{acceptance date}
\published{published date}
\submitjournal{AASJournal name}
\begin{abstract}
Infrared (IR) observations of core collapse supernovae (CCSNe) have been used to infer the mass of dust that has formed in their ejecta. A plot of inferred dust masses versus supernova (SN) ages shows a trend of increasing dust mass with time, spanning a few decades of observations. This trend has been interpreted as  evidence for the slow and gradual formation of dust in CCSNe. Observationally,  the trend exhibits a $ t^2$ behavior,  exactly what is expected from an expanding optically thick ejecta.  In this case, the observed dust  resides in the infrared (IR)-thin ``photosphere" of the ejecta, and constitutes only a fraction of the total dust mass. We therefore propose that dust formation proceeds very rapidly, condensing most available refractory elements within two years after the explosion.  At early epochs, only a fraction of the dust emission escapes the ejecta accounting for the low observed dust mass. The ejecta's entire dust content is unveiled only a few decades after the explosion, with the gradual decrease in its IR opacity.  Corroborating evidence for this picture includes the early depletions of refractory elements in the ejecta of SN1987A and the appearance of a silicate emission band around day 300 in SN~2004et. \end{abstract}
\keywords {ISM: interstellar dust -- Stars: nucleosynthesis -- Stars: supernovae -- Infrared: general, ISM}

\section{INTRODUCTION}
CCSNe drive the chemical enrichment of galaxies and are potentially the most important source of interstellar dust \citep[][and references therein]{dwek06}. The dust formation efficiency depends on a multitude of factors which determine the abundance of refractory elements, and the evolution of the gas density, temperature, and ionization fraction in the ejecta.  Calculating the yield of dust in SN ejecta poses a particular theoretical challenge because of the presence  of radioactive nuclei that generate a cascade of high energy photons and non-thermal electrons which have an important effect on the chemical reaction and nucleation rates in the ejecta. Numerous models have been developed to calculate the production of dust in CCSNe \citep{clayton79,todini01,bianchi07,nozawa13b,cherchneff13,sarangi15,sluder18}. A common result of all models is  that about half the final dust mass is formed within 2 to 3 years after the explosion. The need for the early formation of dust is a natural consequence of the rapid decrease in the density and temperature of the ejecta, which leads to prohibitively long timescales for the nucleation and growth of grains at late times. 

The case for the early and rapid dust formation is supported by the early ``disappearance" of refractory elements, such as Si, Mg, and Fe, from the ejecta of SN~1987A \citep{danziger91,lucy91, dwek92c}. Although these disappearances could have been the results of their recombination, their concurrence with the appearance of the IR excess in the SN light curve strongly suggests dust formation as the cause. 

Circumstantial evidence for the need of early dust formation arises from the high temperatures required to form silicates. They are an important dust component in SN ejecta. In the interstellar medium (ISM) silicates are identified by the 9.7 and 18~\mic\  features that arise from the Si-O stretching and O-Si-O bending modes  in the SiO$_4$ tetrahedral structure. The formation of this structure requires temperatures in excess of $\sim 1000$~K for a period of $\sim 1-2$~days \citep{hallenbeck98}. These conditions are only attainable in SN ejecta during the first few years after the explosion. 

Nonetheless, many researchers believe that dust formation is a slow process, and  that most of the dust is formed at low temperatures, around 10--20~yr after the explosion.  
This scenario is founded on the widely cited Figure~4 in \cite{gall14}, which shows a trend, hereafter referred to as the Mdust--Age trend, of increasing dust mass with SN age. In this figure, dust masses increase  from a value of about $10^{-5}$~\msun\ at about a few years after the explosion to about 0.5~\msun\ after a few decades.  The latter value corresponds to the inferred dust mass in ejecta of SN1987A \citep{matsuura15} on day $\sim$8500 after the explosion.  However, the figure includes mass estimates from several Type IIn supernovae (SN) whose emission is dominated by IR echoes from pre-existing dust or emission from dust that was formed by the interaction of the SN shock wave with the ambient circumstellar medium \citep[][Dwek et al. 2019, in preparation]{andrews11, sarangi18}. In spite of the diverse sources of IR emission, the Mdust--Age trend  has been  widely interpreted  as evidence for the slow and gradual formation of dust in SN ejecta \citep{wesson15,krafton17,bevan16,bevan18,gall18}.

In particular,  \cite{wesson15} and \cite{bevan16} used the IR observations of SN~1987A to argue that dust formation in its ejecta proceeded slowly. This interpretation seems to be supported by the slow evolution of the asymmetry in the profiles of emission lines from the ejecta \citep{bevan16,bevan18}. This asymmetry, manifested in the absorption of the red wings in the lines' profile, is commonly attributed to intervening absorption by newly-forming dust in the ejecta \citep[e.g.][]{lucy89,lucy91,gall14}. Using the slow evolution of the line asymmetries in SN~1987A, \cite{bevan16} concluded that only about 20\% of the dust mass has formed by  day 4000. 
%
%

 In this paper we show that the apparent Mdust--Age trend actually reflects the evolutionary trend of the ejecta's IR opacity. This point can be illustrated by a simple playback of the observed dynamics of SN~1987A \citep{dwek15}. Expanding at a spherically-averaged velocity of $\sim$~900~\kms\ \citep{indebetouw14}, the 200~\mic\ opacity generated by 0.5~\msun\ of ejecta dust at the age of $\approx$ 20~yr is about 0.4. This transparency allowed for the determination of dust mass at that epoch. However, during the early epochs of mid-IR observations \citep{bouchet91a,dwek92c,wooden93}, the same amount of dust in the ejecta would have a  20~\mic\ opacity of about 9100 and 5700 on days 615 and 775, respectively \citep{dwek15}. A dust mass of 0.5~\msun\ could therefore have easily been ``hidden" in the optically thick ejecta at those early epochs. These conditions apply not only to SN~1987A, but to CCSNe in general, necessitating the reevaluation of the evolution of dust mass in SN ejecta.

The paper is organized as follows. In Section~2 we develop a simple model for the evolution of the ejecta opacity, showing that it can offers a natural explanation for the observed Mdust--Age trend. In Section~3 we present corroborative arguments for the rapid formation of dust in the ejecta.  A brief summary of the paper is presented in Section~4              

 \section{The evolution of ejecta opacity}
 
The  dust mass, $M_d^{obs}$, is inferred from the observed specific IR flux, $F_{\nu}^{obs}(\lambda)$, using the relation
\begin{equation}
\label{md}
M_d^{obs} = {F_{\nu}^{obs}(\lambda)\ D^2\over B_{\nu}(\lambda, T_d)\, \kappa(\lambda)}
\end{equation}
where $D$ is the distance to the source, $\kappa(\lambda)$ is the dust mass absorption coefficient, and $B_{\nu}(\lambda, T_d)$ is the Planck function at wavelength $\lambda$ and dust temperature $T_d$.
Equation~(\ref{md}) assumes that the source is optically thin and all the dust is ``visible'' to the observer. 

The radial opacity, $\tau(\lambda, t)$ of a spherically-expanding expanding  dusty ejecta at wavelength $\lambda$ and time $t$ is given by
\begin{eqnarray}
\label{tau}
\tau(\lambda, t) & = & \rho_d(t)\, R(t)\, \kappa(\lambda) = {3\over 4}\, \left({M_d(t) \over \pi R(t)^2}\right) \kappa(\lambda) \nonumber \\
 & = & {3\over 4}\, \left({M_d(t)\over \pi v_{ej}^2}\right) \kappa(\lambda)\, t^{-2} 
\end{eqnarray}
where  $\rho_d(t)$ and  $M_d(t)$ are, respectively,  the mass density and total mass of dust in the ejecta at time $t$,   and $R(t)$ is the radius reached by a freely expanding ejecta at a velocity $v_{ej}$ at $t$.

The escape probability of an IR photon from a sphere of optical depth $\tau$ is given by \cite{cox69,osterbrock06}
\begin{equation}  
\label{pesc}
P_{esc}(\tau) = {3\over 4 \tau}\ \left[1-{1\over2\tau^2}+({1\over\tau}+{1\over2\tau^2})e^{-2\tau}\right]
\end{equation}
where the $\lambda$ dependence of $\tau$ has been suppressed for sake of clarity.
At large optical depths $P_{esc}(\tau) \approx 3/4\tau$. It approaches a value of $\sim \exp(-3\tau/4)$ for $\tau << 1$.  

\begin{figure}
\begin{center}
\includegraphics[width=3.3in]{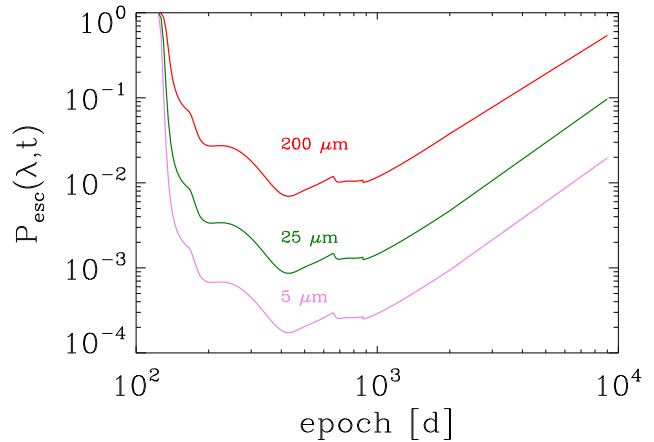}
\caption{\label{pescf} \footnotesize{The evolution of the escape probability,  given by eq.~(\ref{pesc}), at different wavelengths.  }}
\end{center}
\end{figure}

\begin{figure}
\begin{center}
\includegraphics[width=3.3in]{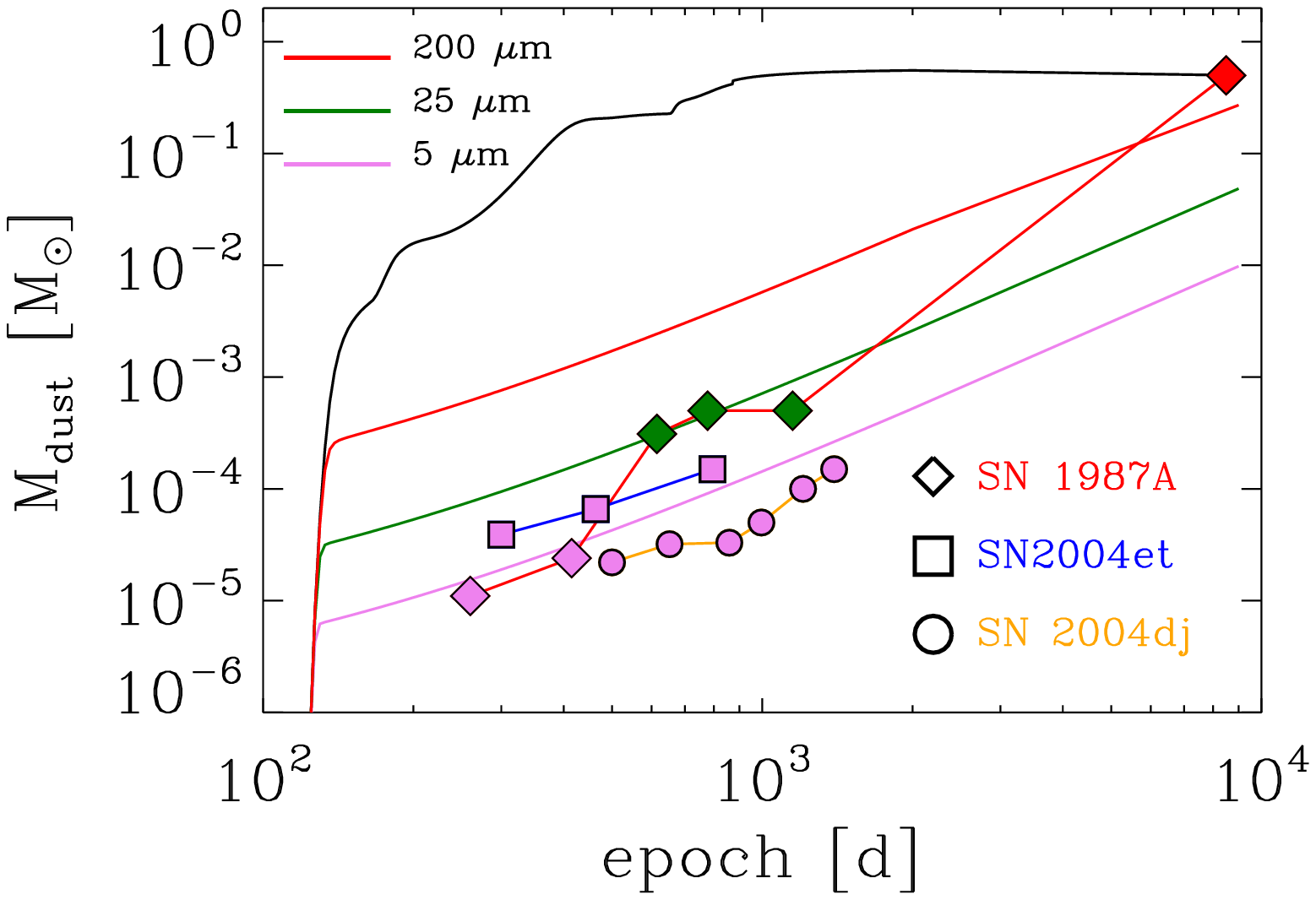}
\caption{\label{mdust} \footnotesize{The growth of the dust mass in the ejecta of a 19~\msun\ SN (black curve) is plotted against the epoch of observations, measured since the time of the explosion \citep{sarangi15}. The symbols represent the observationally inferred dust masses for various SNe that were observed for at least three epochs: SN~1987A \citep{wooden93,dwek92c,matsuura15} (diamonds); SN~2004et \citep{kotak09} (squares); SN~2004dj \citep{kotak05, szalai11,meikle11} (circles). The different colored curves show the evolution of the inferred dust mass with ejecta opacity at three different wavelengths of observations.  Observed dust masses were color coded according to the wavelength of observations. }}
\end{center}
\end{figure}

Figure~\ref{pescf} shows the evolution of the escape probability at several wavelengths. In calculating $\tau$, the mass absorption coefficient was represented by a $\lambda^{-1}$ power law, normalized to a value of $10^3$~cm~g$^{-1}$ at 5~\mic. The expansion velocity of the ejecta was take to be of $10^3$~\kms, approximately equal to the spherically-averaged expansion velocity of the ejecta of SN~1987A \citep{indebetouw14}. The evolution of the ejecta dust mass was taken from \cite{sarangi15}. It is shown in Figure~\ref{mdust}, and described in more detail below. 
Figure~\ref{pescf} shows that the ejecta is transparent at the onset of dust formation, but becomes rapidly  opaque after day $\sim 150$.

The observed flux, $F_{\nu}^{obs}(\lambda)$, is  given by $F_{\nu}^{obs}(\lambda) = F_{\nu}^{0}(\lambda)\, P_{esc}(\tau)$, where $F_{\nu}^{0}(\lambda)$ is the unobscured flux from the ejecta. The  observed dust mass at time $t$ observed at a given wavelength $\lambda$, $M_d^{obs}(\lambda, t)$, derived from the optically thin assumption, is related to the total dust mass by
\begin{eqnarray}
\label{mvis1}
M_d^{obs}(\lambda, t)  & =  & {F_{\nu}^0(\lambda)\,  P_{esc}(\tau)\ D^2\over  B_{\nu}(\lambda, T_d)\, \kappa(\lambda)} \nonumber \\
 & =  & M_d(t)\times P_{esc}(\tau)  
 \end{eqnarray} 
 
 Equation~(\ref{mvis1}) illustrates the degeneracy between the dust mass in the ejecta and the optical depth. Different combinations of $M_d$ and $P_{esc}(\tau)$ can yield the same observed dust mass. Previous studies that concluded that the ejecta is optically thin \citep[e.g][]{wooden93, meikle11} made this assumption implicitely in their calculations. 

When the ejecta is optically thick, $P_{esc}(\tau) = 3/4\tau$, and substituting the expression for $\tau$ from eq. (\ref{tau}), equation~(\ref{mvis1}) can be written as
\begin{eqnarray}
\label{mvis2}
M_d^{obs}(\lambda, t)  & = & {3M_d\over 4 \tau} = {\pi\, R(t)^2\over \kappa(\lambda)}  \nonumber \\
& = & \pi \rho_d(t)\, R(t)^2 h(t) =  \left[{\pi\, v_{ej}^2\over \kappa(\lambda)}\right] \, t^2 
\end{eqnarray} 
where $h\equiv (\rho_d\, \kappa)^{-1}$ is the depth at which the ejecta's atmosphere reaches a value of $\tau = 1$.
Equation~(\ref{mvis2}) shows that at high optical depths the ``visible" dust mass is independent of the actual dust mass, and that it increases as $t^2$.

Figure~\ref{mdust} (black curve), taken from \cite{sarangi15}, depicts the growth of the dust mass in the ejecta of  a 19~\msun\ progenitor star calculated with a detailed chemical reaction/dust formation network  \citep{cherchneff09,cherchneff10b,sarangi13}. The calculated dust mass was normalized to reproduce the inferred dust mass of SN~1987A at late epochs. A similar evolutionary profile for the evolution of dust mass in SN~1987A was recently derived by \cite{sluder18}. Figure~\ref{mdust} represent the dust mass, given by $M_d(t)\times P_{esc}(\tau)$. The escape probability is wavelength dependent, and the different colored lines depict the evolution of $M_d^{obs}$  at different wavelengths. The figure shows that at each wavelength, $M_d^{obs}(t)$ rapidly approaches the  $t^2$ behavior given by equation~(\ref{mvis2}) for $\tau \gg 1$.  

The colored symbols represent the inferred dust masses for several SNe. The IR spectrum of SN~1987A  was observed around the peak wavelength of the emission, at about 5~\mic\ on days 260 and 415, $\sim 10$~\mic\ on days 615 and 775,  $\sim 25$~\mic\ on day 1140, and at $\sim 200$~\mic\ on day 9000. The inferred dust masses are in very good agreement with those expected at the observed wavelengths. 
SN~2004et and SN~2004dj were observed with the IR Array Camera (IRAC) instrument on board the {\it Spitzer} spacecraft. The 3.6-8~\mic\ emission is dominated by ejecta dust, whereas the emission at longer wavelength emission may be an IR echo from the SN UV-optical light curve \citep{meikle11}. 
The figure shows that the inferred dust masses from these SNe agree well with the calculated visible dust mass at 5~\mic, roughly the peak of the ejecta dust emission. 

Overall, the figure  shows a good agreement between the inferred dust mass and that predicted from eq.~(\ref{mvis2}) at the observed wavelengths. In particular, the data follow the predicted $t^2$ behavior. 
Offsets of different SNe from the theoretical Mdust-age curve along the y-axis reflect the
   different expansion velocities and different dust compositions, leading to different values of $\kappa$.
      
In \cite{gall14}, the evolution in the ejecta dust mass was approximated by a broken power law. However, data prior to day $\sim$~240, the position of the break in the power law, cannot be attributed to newly-formed ejecta dust. Calculations \citep[e.g.][]{sarangi15,sluder18} and observations of SN~1987A \citep{wooden93} show that dust formation commences only after day $\sim 250$, when the ejecta has cooled sufficiently. Data prior to this epoch is affected by emission caused by an echo from pre-existing circumstellar dust \citep[e.g.][]{bode80b,dwek83b}. Data beyond this epoch are not well sampled, except for SN~1987A which, as shown in Figure~\ref{mdust},  clearly exhibits a $t^2$ behavior.

For clarity, all calculations presented above  assumed a homogeneous ejecta expanding at a constant velocity. 3D hydrodynamic simulations of SN explosions show that instabilities generated by the expanding radioactive Ni-Co bubble  will cause the ejecta to fragment into clumps \citep[][and references therein]{wongwathanarat15}. The early escape of gamma- and X-ray emission from SN~1987A  provided observational evidence for large-scale instabilities and mixing in the  ejecta \citep[see review by][]{mccray16}.  The IR emission from a clumpy ejecta was discussed in detail by \cite{dwek15}. When the clump filling factor remains unchanged, a clumpy ejecta will  exhibit the same trend of evolving opacity as a homogeneous one, as each expanding clump becomes optically thin. The $t^2$ behavior of the Mdust-Age trend exhibited in this paper will thus remain unchanged for a clumpy ejecta.

So far we attributed the low inferred dust mass to the IR opacity of the ejecta. We illustrated this point in \cite{dwek15} by assuming that the silicate and carbon grains each have distinct single temperatures. Their emission peaks at around the wavelengths of observations, so that all dust is, in principle, equally observable. Alternatively, the low inferred dust mass could reflect the fact that most of the dust in the ejecta is cold, and therefore not observable with the mid-IR wavelengths used at the early epochs of observations. Such possibility will give further support for the early and rapid formation of dust in the ejecta. However, modeling such scenario will require detailed knowledge of the relative spatial distribution of the radioactivities and the dust, and is beyond the scope of the present paper.  

 \section{The growth of dust mass in SN ejecta}
IR observations of SN~1987A show the appearance of an IR emission component, in excess of the photospheric emission, within a year of the explosion \citep{wooden93}. The appearance of the IR excess coincided with the sudden drop in the Mg~I]~0.4571\mic\ and [Si~I]~1.65\mic\ emission lines around day 530  \citep{danziger91, lucy91}, and a drop in the bolometric luminosity of the SN \citep{whitelock89}. These observations strongly suggest that  the IR emission originated from silicate dust that formed rapidly and efficiently in the ejecta, and that caused the partial obscuration of the UVO output from the SN. The absence of the 9.7 and 18~\mic\ silicate features in the spectra is the result of self absorption in the optically thick ejecta in the model of \cite{dwek15}. Alternatively, \cite{wesson15} argued that the absence of the silicate features is caused by the dominance  of the emission from featureless carbon dust, which will require carbon to be the dominant dust species formed in the ejecta, contrary to theorretical model predictions \citep{sarangi15,sluder18}.  

The early appearance of a broad emission feature in the 8--14~\mic\ spectrum of SN~2004et on day 300 also provides evidence for the early formation of dust in SN ejecta \citep{kotak09}. The feature, and its related photometric 8~\mic\ excess over that of a blackbody spectrum, persisted through day 690 and faded thereafter. The feature is attributed to silicate dust emission, and its decline could be the results of a drop in dust temperature. Observations of SN~1987A show that after day 600 the dust temperature has dropped below $\sim 400$~K \citep{wooden93}, the temperature at which the emission peaks around 8~\mic. The drop in dust temperature could  result from the decline in the radioactive heating source or the growth of the dust grains. The decline of the feature could also have been caused by the growth of the silicate grains to a radius of a few microns, above which they become opaque to their own feature.  

The early and rapid formation of dust in SN ejecta is also widely supported by theoretical models for the formation of dust in SN ejecta \citep[e.g.][]{todini01, nozawa10, cherchneff10b, sarangi15}. Most recent molecular nucleation models for the formation of dust in SN~1987A show that most of the dust formed before day 1000 after the explosion \citep[][Figure 10 in their paper]{sluder18}. Their computed dust formation rate agrees with the early dust formation model, and does not support the gradual formation scenario of \cite{wesson15} and \cite{bevan16}.
 
The strongest argument for the slow formation of dust in the ejecta of SN~1987A is the evolution of the ejecta line asymmetries \citep{bevan16}. However, their model adopts a spherically symmetric expanding ejecta, whereas resolved 2012 observations of the SN with the Atacama Large Millimeter/Submillimeter Array (ALMA) suggests a clumpy elongated structure \citep{indebetouw14}. Late time observations (day 10,000) of the  H$\alpha$ and the blended [Si~I]+[Fe~II] line emission from SN~1987A were used to construct a 3D model for the distribution of the emitting material \citep{larsson16}. The observations show that the blended 1.644~\mic\ line is asymmetric with a prominent redshifted component, which is inconsistent with the model predictions of  \cite{bevan16}. The 3D distribution of the [Si~I]+[Fe~II] line reveals an elongated structure that is consistent with the ALMA observations, but with a gap at the center, which is not detectable in the ALMA image due to projection effects. These results suggests that the models of \cite{bevan16} and \cite{bevan18} are at least incomplete, and not robust enough to corroborate a scenario for the slow and gradual formation of dust in the ejecta. A more complex model consisting of a clumpy asymmetrically expanding ejecta is required to obtain a consistent picture for the evolution of the line profiles from the ejecta.
 
The detection of a $^{49}$Ti excess in presolar SiC grains of SN origin, commonly referred to as X-grains, was interpreted by as evidence for the late formation of silicate carbide in type~II SNe \citep{liu18}. 
Detailed calculations show that $^{49}$Ti must have been incorporated in the grains not sooner than 2 years after the explosion \citep{liu18}, in agreement with theoretical calculations showing that SiC grains are late to form in the ejecta \citep{sarangi15}.
Figure~\ref{mdust} shows that at that epoch more than 50\% of the dust has already formed in the ejecta. The mass of SiC dust grains is about $10^{-4}$~\msun, and the mass of all Ti isotopes is about $10^{-4}$~\msun\ \citep{nomoto13}. The X-grains make therefore a negligible contribution to the total dust mass in the ejecta. The $^{49}$Ti anomaly is therefore an excellent chronometer for the formation time of the X-grains, but has no bearing on the epoch of the bulk of the dust formation in the SN ejecta.

 \section{Summary}

In this paper we have offered a natural explanation for the Mdust--Age trend for CCSNe. The observable dust mass constitutes only a small fraction of the total mass, namely that present in the ``photosphere" of the IR-thick ejecta. As the ejecta expands its optical depth decreases as $R(t)^{-2} \sim t^{-2}$, consistent with the observed Mdust--Age trend. We therefore propose an alternate scenario in which most of the dust forms within two year after the explosion. The IR emission from the dust is initially self absorbed, revealing most of the dust mass only decades after the expansion \citep{dwek15}. 
The rapid formation of dust in the ejecta is consistent with the observed depletion of refractory elements concomitant with the early dust emission and the drop in the bolometric luminosity from some SNe. It is also consistent with theoretical models, in which most of the dust forms during the early phases of the expansion when ejecta densities and temperatures are high enough to form the necessary chemical bonds and support the reaction rates needed to grow the nucleation seeds.

Our scenario also offers a simple solution to the conflict between theoretical calculations that show the dominance of silicate dust over carbon \citep[e.g.][]{sluder18}, and the models of \cite{wesson15} and \cite{bevan16} in which carbon is the dominant dust species. In our scenario, the lack of the 9.7 and 18~\mic\ silicate emission features in the spectra of SN~1987A is not evidence for the absence of silicate dust, but merely a manifestation of the optical depth of the ejecta \citep{dwek15}.

The results of this paper have important implications for determining the physical conditions in SN ejecta. Modeling the UVO-NIR emission from SN~1987A requires calculations of the energy cascade that degrades  the $\gamma$-rays and positrons into UV and optical and IR line emission. These detailed calculations were done with the assumption that the dust opacity is low \citep{jerkstrand11}. However, because of the significantly large UVO optical depth,  the competition between the dust and atomic species for the absorption of these photons must now be taken into account.

Dedicated ground-based or satellite (e.g. the {\it James Webb Space Telescope, JWST}) observations of young SNe  in local galaxies at $z \lesssim 0.2$, over a period of several years will provide the much needed observations to confirm the $t^2$ trend in Mdust-Age relation.

\acknowledgements
This work was supported by NASA's 16-ATP16-0004 research grant. We thank the anonymous referee for his/her thorough reading of the manuscript and constructive comments that led to  important improvements in the manuscript. We also thank  Ori Fox and Larry Nittler for helpful discussions.

\clearpage
\bibliographystyle{$HOME/Library/texmf/tex/latex/misc/aastex52/aas.bst}
\bibliography{$HOME/Dropbox/science/00-Bib_Desk/Astro_BIB.bib}

 \end{document}